\documentclass{jfm}
\usepackage{graphicx}
\usepackage{epstopdf, epsfig}
\usepackage{amsmath}
\usepackage{color}
\usepackage{bm}
\usepackage{mathrsfs}
\usepackage{subfigure}
\usepackage{caption}
\usepackage{booktabs}

\newcommand{\ue}{\mathrm{e}}
\newcommand{\ud}{\mathrm{d}}

\shorttitle{Granular dynamics in auger sampling}
\shortauthor{Y. Feng, and others}

\title{Granular dynamics in auger sampling}

\author{Yajie Feng\aff{1},Shuo Huang\aff{1},Yong Pang\aff{2},
Kai Huang\aff{3,4}
\corresp{\email{kh380@duke.edu}},
\and Caishan Liu\aff{1}
\corresp{\email{liucs@pku.edu.cn}}}

\affiliation{\aff{1}State Key Laboratory of Turbulence and Complex System, College of Engineering, Peking University, Beijing, 100871, China 
\aff{2}Beijing Spacecrafts, China Academy of Space Technology, Beijing, 100090, China 
\aff{3} Division of Natural and Applied Sciences, Duke Kunshan University, 215306 Kunshan, Jiangsu, China
\aff{4} Experimentalphysik V, Universit\"at Bayreuth, 95440 Bayreuth, Germany
}

\begin{document}

\maketitle

\begin{abstract}

From geotechnical applications to space exploration, auger drilling is often used as a standard tool for soil sample collection, instrument installation, and  others. Focusing on granular flow associated with the rotary drilling process, we investigate the performance of auger drilling in terms of sampling efficiency, defined as the mass ratio of the soil sample collected in the coring tube to its total volume at a given penetration depth, by means of experiments, numerical simulations, as well as theoretical analysis. The ratio of rotation to penetration speed is found to play a crucial role in the sampling process. A continuum model for the coupled granular flow in both coring and discharging channels is proposed to elucidate the physical mechanism behind the sampling process. Supported by a comparison to experimental results, the continuum model provides a practical way to predict the performance of auger drilling. Further analysis reveals that the drilling process approaches a steady state with constant granular flow speeds in both channels. In the steady state, sampling efficiency decreases linearly with the growth of the rotation to penetration speed ratio, which can be well captured by the analytical solution of the model. The analytical solution also suggests that the sampling efficiency is independent of gravity in the steady state, which has profound implications for extraterrestrial sample collection in future space missions. 

\end{abstract}


\section{Introduction}\label{sec:introduction}

Granular materials can be considered as complex fluids with finite yield stress that is associated with the transition between solid- and liquid-like states~\citep{Jaeger1996,Andreotti2013}. Because of the highly dissipative and heterogeneous nature of granular materials, a generally applicable description of granular materials as continuum is still lacking, despite of continuous efforts in the past decades~\citep{Jenkins1983,Goldhirsch2003,Forterre2008}. Concerning widespread examples of handling granular materials in nature, industrial sectors, and our daily lives~\citep{Duran2000,Aguirre2021}, it is essential to understand the response of granular materials to disturbance by rigid objects such as an auger~\citep{Imole2016}. In this regard, there have been extensive investigations on, for instance, impact of granular jet on a rigid plane~\citep{Mueller2014} or reciprocally projectile impact on granular media~\citep{Colaprete2010,BrzinskiIII2010,Meer2017,Huang2020}, crater formation~\citep{Ruiz2013}, as well as bio-mechanical topics including drag reduction through self-propulsion used by organisms\citep{Liu2011,Jung2017,Texier2017} and locomotion in granular systems~\citep{Aguilar2016}. 

From the application perspective, drilling into granular media by means of helical motion for sample collection, instruments installation or construction purposes finds applications in civil, chemical engineering as well as conventional energy sectors. As such, a wide variety of screw conveyors can be found in chemical and process engineering industries to enhance the transport and mixing of granular materials~\citep{Xiong2015,Pang2018}. In the new era of space exploration, the exploration of extraterrestrial regolith in terms of granular sample collection leads to the deployment of various types of granular samplers for exploring the geological evolution of extraterrestrial bodies, such as the Luna probe project from the former Soviet Union and America's Apollo project~\citep{Zacny2008}. Due to the advantages of auger transport, various drill samplers have been developed for different space exploration projects, such as ESA's MOONBIT project~\citep{poletto2015seismic}, NASA's ExoMars project~\citep{zacny2004laboratory,firstbrook2017experimental} and Japan's LUNAR-A mission~\citep{nagaoka2010experimental,nakajima1996lunar}. China's Chang'e lunar exploration project  also used an auger drill sampler to collect and return subsurface lunar regolith~\citep{zhang2017drilling,quan2017drilling}. Although auger transport has been widely implemented in the applications, modeling auger conveying of granular materials is still a challenging subject~\citep{imole2016reprint}.

In connection to the fundamental understanding of granular drag, continuous investigations have been devoted to auger conveying of granular materials by means of constitutive models~\citep{yu1997theoretical,roberts1999influence,dai2008model}, experiments~\citep{imole2016reprint,waje2006experimental,ramaioli2008granular}, and numerical simulations using either computational fluid dynamics (CFD)~\citep{xiong2015characterizing,duan2017modified} or discrete element methods (DEM)~\citep{ramaioli2008granular,owen2009prediction,shimizu2001three} in the past decades. Most of the studies show that operating conditions, such as the rotational speed of the auger, the inclination of the auger conveyor, and the initial filling fraction of the bulk materials, significantly affect the performance of an auger conveyor. It was found that both the intruder's configuration~\citep{gravish2010force,guillard2014lift} and velocity~\citep{uehara2003low,katsuragi2007unified} significantly affect the drag force. In particular, recent experiments revealed configurations for a rotating cylinder to drill inside granular materials with surprisingly low torque~\citep{guillard2013depth,liu2017forces}. This is in agreement with weakened resistance of soil against penetration of a spinning cone~\citep{jung2017reduction} or a rotating helix~\citep{liu2011force} found in experiments.

More recently, the drill used in China's Chang'e lunar exploration project has been a subject of series investigations, particularly on the interactions between the soil and the auger. \citet{zhang2017drilling} numerically and experimentally investigated the penetration force and rotational torque of the drill, and  found that the penetration force can be reduced due to self-propulsion. \citet{quan2017drilling} proposed an index to characterize the condition for the occurrence of choking, a phenomenon in which the cuttings build up in the auger flight and cause the rotational torque to increase sharply~\citep{statham2012automated}. \citet{Zhao2016Soil} found a maximal removal capability for the cutting  conveying. \citet{Tang2017,tang2018experimental} illustrated the coupling between the granular flow in the auger flight and that in the coring tube. Specifically, these experimental results revealed that the coring results crucially depend on the drilling conditions, characterized by an index called penetration per revolution (PPR). If the PPR value is not suitable, the drill may either enter into a failure mode due to choking or only sample a small amount of soil. For a successful soil-sampling, a proper PPR value should be selected according to the physical properties of the soil.

Here, we focus on auger conveying of granular materials relevant to the drill tool shown in figure~\ref{fig:Config}. The drill tool consists of a drill bit, a helical and right-handed auger, and a hollow coring tube. In the drilling platform, soil is displaced in the following three processes~\citep{zhang2017drilling-1,li2017design}:(i) The cutting process, in which the stiff soil is loosened by the drill bit~\citep{perneder2012bit}; (ii) The discharging process, in which the cuttings are removed from the bottom of the borehole to the ground surface; (iii) The coring process, in which the soil is sampled into the hollow coring tube. Generally speaking, the three processes are coupled together to affect both drilling loads and coring results. This investigation aims at modeling the latter two processes as the efficiency of soil sampling relies predominantly on them.

Experimental results suggest the ratio between penetration and rotation speeds to be an important dimensionless parameter controlling sampling efficiency. Theoretically, a one-dimensional continuum model is established to describe granular flow in both coring and discharging channels. Quantitative comparisons to the experimental results indicate that our model successfully captures the essential role played by the ratio between penetration and rotational speed in determining the coring results. Finally, steady state analysis of granular flow yields an analytical prediction of sampling efficiency, providing a practical way to control sample collection with auger drilling.

The remainder of this article is organized as follows: \S~\ref{sec:experiment} briefly introduces the experimental setup and presents the experimental results obtained from different drilling conditions and different types of soils. The governing equations describing granular flows in the coring and discharging channels are developed in \S~\ref{sec:theory}. We compare experimental results with numerical ones from the theoretical model in \S~\ref{sec:validation} and analyze the steady-state solution of the model in \S~\ref{sec:steady}. Finally, we conclude with an outlook for further investigations in \S~\ref{sec:conclusion}.

\section{Experiment}\label{sec:experiment}
\begin{figure}
  \centerline{\includegraphics[width=5.0in]{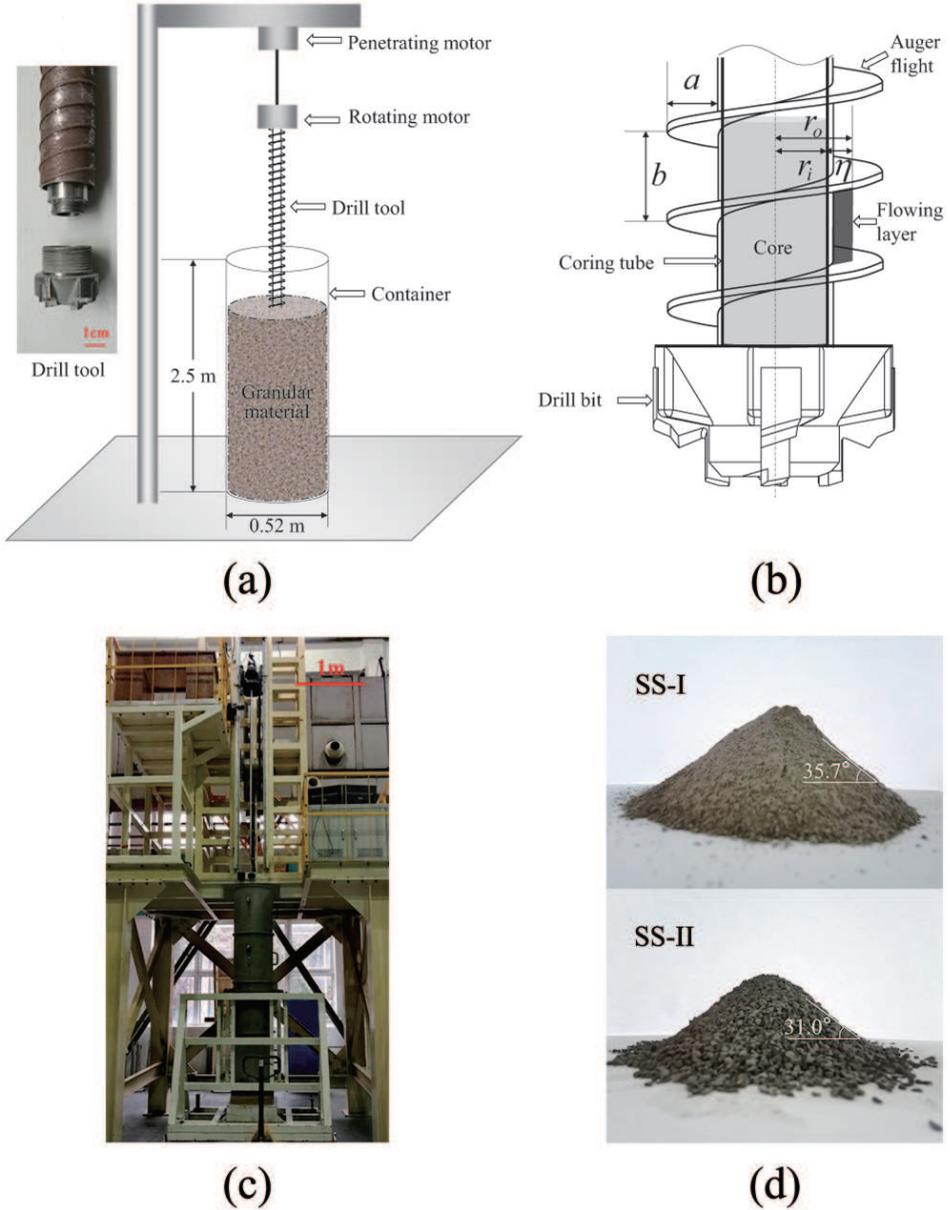}}
  \caption{ Schematic of the experimental apparatus (\textit{a}) and the drill tool (\textit{b}) with geometrical variables defined. Inset of (a) is a snapshot of the drill bit detached from the coring tube used in the experiments. (c) and (d) correspond to the experimental set-up and snapshots of the two lunar simulants with their angles of repose marked. Note that plot (\textit{b}) is not to scale.}\label{fig:Config}
\end{figure}

As illustrated in figure~\ref{fig:Config}, the experimental apparatus mainly consists of a drill platform, a drill bit, a helical and right-handed hollow auger, a sampling device and a soil container. The sampling device contains a coring tube inside the hollow auger. The cylindrical container has a diameter and height of $0.52$\,m and $2.5$\,m, respectively. The geometric profile of the auger can be defined by four parameters: Auger flight radius $r_o=1.75$\,cm, coring tube radius $r_i=1.55$\,cm, pitch $b=1.20$\,cm, blade thickness $t_c=0.10$\,cm and groove depth $a=r_o-r_i=0.20$\,cm. For sample collection, a soft bag is attached to the inner surface of the coring tube. Throughout the entire drilling process, the coring tube, together with the sample collected, moves along with the auger without rotation.

Before drilling starts, the granular sample is compacted by vibration to create a reproducible initial condition. More specifically, we incorporate a five-stage sample filling and vibration process to ensure a dense initial packing. Based on the maximum packing density of a specific sample, we add each time $1/5$ of the total mass (note that about one ton of sample is used in each experiment) into the container. Initially, the whole container is vibrated in the vertical direction against gravity at $30$ Hz for $5$ minutes. Subsequently, tri-axial vibrations are applied at the same frequency for $20$ minutes to further compact the sample. Based on a previous investigation~\cite{Nowak1998}, the number of taps through this process (close to $10^5$) is sufficient for the system to reach a steady state. Finally, the height of the granular layer is monitored to check whether the desirable packing density is achieved or not. If not, $10$ minute tri-axial vibrations are applied additionaly to compact the sample, before the whole process repeats for the next batch of sample.

Subsequently, the drill tool rotates and penetrates synchronously into the granular sample. The initially compacted soil surrounding the drill tool is then fluidized as the drill bit cuts through. As shown in Fig. \,\ref{fig:Config}(a), the drill bit includes four cutting edges organized symmetrically about central axis. The diameter of the inner tube matches that of the coring tube to facilitate sample flow from the drill bit to the coring tube. The outer radius of the drill bit is slightly larger than that of the auger flight for effectively fluidized lunar simulants to flow through the outer channel. After fluidization, the soil is transported upwards through either the coring tube or the auger. Once the target depth is reached, both penetration and rotational motions stop simultaneously, meanwhile the coring tube is closed by a sealing device to complete the sample collection process.

Experiments have shown that the coring results are determined by the physical properties of the soil and the  kinematic parameters of a drill tool~\citep{zhang2017drilling,Tang2017,tang2018experimental}. In the experiment, we use two types of simulated lunar soils~\citep{Carrier2003particle} with gray basaltic pozzuolana as the main component. The grain size in the simulated Soil-I (SS-I) ranges from 0.1 to 1 mm, and that in the simulated Soil-II (SS-II) ranges from 1 to 2 mm. The bulk densities of the two simulated soils are $\rho_{\text{I}}= 2.13$ g/cm$^3$ and $\rho_{\text{II}}=1.85$ g/cm$^3$, respectively. The packing fractions of the two soils are $\psi_{\text{I}}=0.71$ and $\psi_{\text{II}}=0.60$. As shown in Fig.\,\ref{fig:Config}(d), their internal friction angles are $\phi_{\text{I}}=35.7^\circ$ and $\phi_{\text{II}}=31.0^\circ$, respectively.

\begin{figure}
  \centerline{\includegraphics[width=2.8in]{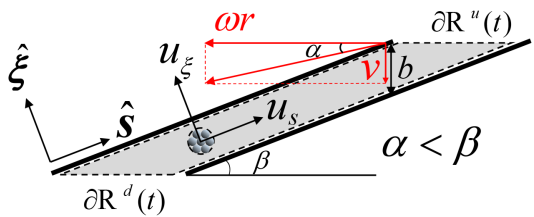}}
  \caption{Granular flow in a segment of the external channel with two boundaries (dashed line). $u_s$ and $u_\xi$ are the components of the absolute flow velocity along $\hat{\bm{s}}$ and $\hat{\bm{\xi}}$ directions, respectively. $v$ and $\omega r$ are the  penetration and rotation velocities for the  point on the auger at radius $r$, and $\alpha \def \arctan{v/(\omega r)}$.}\label{fig:outvela}
\end{figure}

The motion of the drill is determined by two parameters: penetration speed $v$ and rotational speed $\omega$. Feedback loop in motor control is employed to ensure constant $v$ and $\omega$ throughout the drilling process. The experiments are conducted under three different rotational speeds: $\omega =80, 120, 160$\,rpm. For each $\omega$, penetration speed ranges from $10$ to $360$\,mm/min. The target depth is set to $H=1.0$\,m for all experiments.

In order to characterize the geometry and kinematics of the auger flights, we introduce the geometry dependent helical angle $\beta$ and the elevation angle $\alpha$ of the velocity vector for a point $P$ on the auger flight. As shown in Fig.\,\ref{fig:outvela}, the geometry dependent helical angle $\beta$ is defined as

\begin{equation}\label{eq:L1}
 \tan\beta = \frac{b}{ 2\pi r}
\end{equation}
with $r$ the distance of $P$ to the rotation axis. Once drilling starts, $P$ undergoes a helical motion with fixed radial distance and an elevation angle $\alpha$, which can be estimated with 

\begin{equation}\label{eq:L2}
\tan\alpha = \frac{v}{\omega r}.
\end{equation}

Note that, for the special case of $\alpha=\beta$, point $P$ moves along the streamwise direction (flight direction), reminiscent to inserting a straight hollow tube into the granular sample. In this case, the grains in the auger remain static and cannot be discharged. Consequently, the sample height in the coring tube equals the drilling depth. If $\beta < \alpha$, the drill drives granular particles downward and the enforced compaction may lead to chocking at the bottom of the drill. When $\beta > \alpha$, the drill drives the sample upward. As such, the relation between $\alpha$ and $\beta$ is crucial in the drilling process. Thus, we define speed ratio $\gamma$ as a control parameter:
\begin{equation}\label{eq:L3}
\gamma = \frac{\tan\beta}{\tan\alpha} = \frac{\omega b}{2\pi v}.
\end{equation}

To quantify the sampling efficiency, we define another dimensionless number
\begin{equation}\label{eq:samplingrate1}
 \zeta=\frac{m_i}{m_{max}},
\end{equation}
where $m_i$ denotes the mass of the sampled soil in the coring tube as it reaches target depth $H$, and $m_{\rm{max}}= \pi \rho H r^2_i$ corresponds to the maximum mass of the soil at $H$ with $\rho$ bulk density of the sample.

\begin{figure}
  \centerline{\includegraphics[width=2.8in]{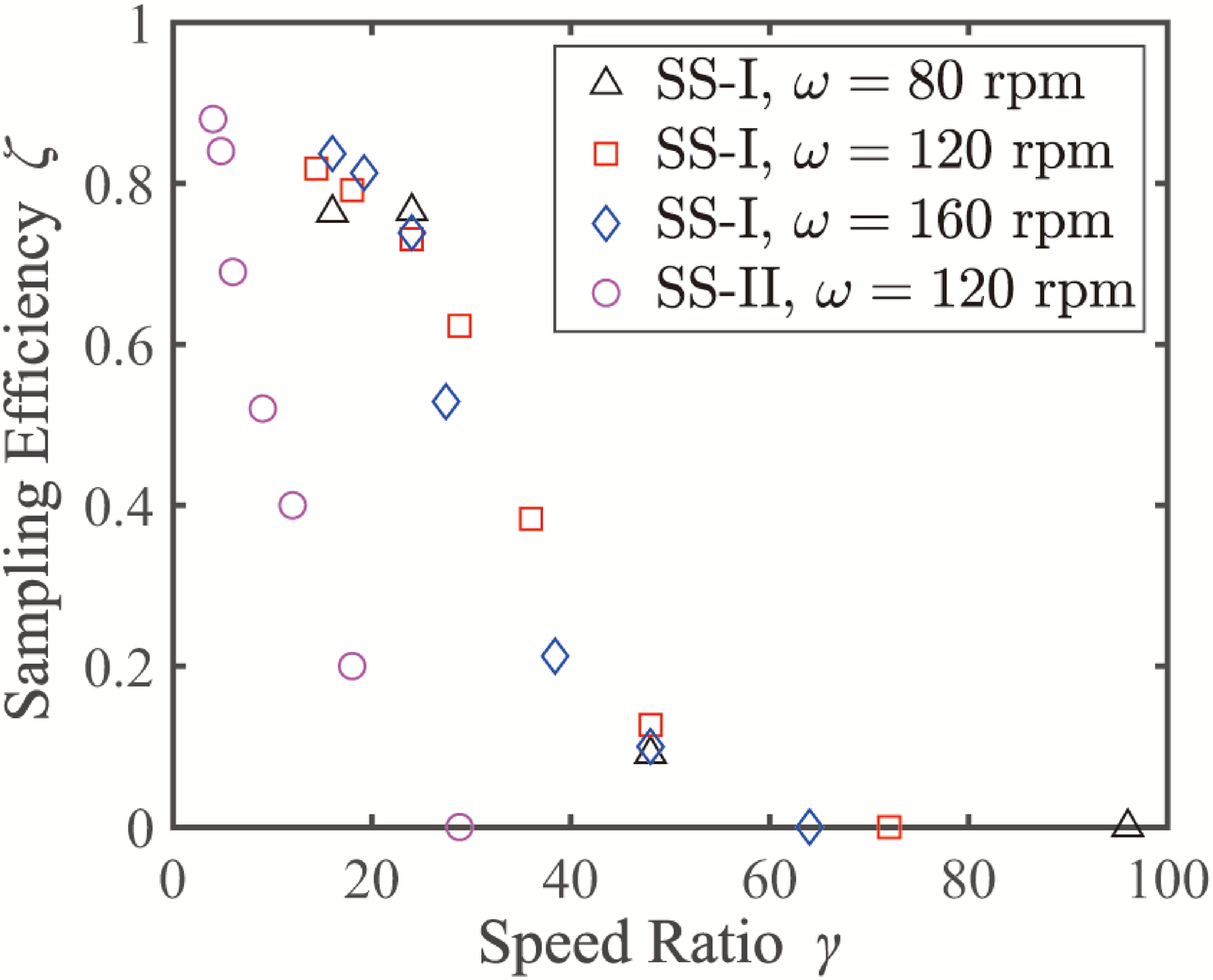}}
  \caption{Sampling efficiency as a function of speed ratio for two different types of soil used in experiments. For Soil-I, there are three
  different rotational speeds of 80, 120 and 160 rpm. For Soil-II, there is only one rotational speed of 120 rpm. We use the same marker to represent the experimental data collected at the same $\omega$ but different $v$. Based on initial test runs, uncertainty of the sampling efficiency is $\sim 10$\%.}
  \label{fig:expsample}
\end{figure}

Figure~\ref{fig:expsample} shows the relations between $\zeta$ and $\gamma$ for two types of soils under different configurations. It shows that: (i) For both types of soils, the sampling efficiency decreases monotonically to $0$ as $\gamma$ grows. (ii) For sample II, a systematic variation of driving conditions yields a master $\zeta-\gamma$ curve. (iii) Each sample type has its own $\zeta-\gamma$ curve. The experimental results suggest that the drilling process is determined by both properties of the granular sample and speed ratio $\gamma$. Note that the lower bound of $\gamma$ is higher for SS-I in comparison to SS-II. This is because highly compacted granular sample with smaller particle sizes requires higher torque to drill into than that with larger particle sizes, particularly for small $\omega$.

\section{Continuum model}\label{sec:theory}

In this section, we introduce a continuum model for granular flow in both internal (in the coring tube) and external (on the auger flight) channels, in order to shed light on the experimental results presented above. As granular sample has to fluidize before being displaced, it can be considered as a fluid. Because its flow in either internal or external channel is confined to a either vertical or helical direction, we consider the sample collection process as one-dimensional flows of incompressible fluids. The packing density change during the fluidization process at the drill bit is not considered here, because the model describes the flow of granular fluids in both internal and external channels. In the future, further experimental analysis on the change of packing density during the initial fluidization process is needed to incorporate compressibility of the granular sample in the model.

In the subsequent parts of the section, we introduce governing equations based on mass and momentum balance for both internal and external channels, as well as the coupling in between. Finally, we conclude with a summary of five governing equations to numerically solve for the time-dependent mass and velocity in both channels as well as the pressure at the bottom of the drill.

\subsection{Flow dynamics in internal channel}
\begin{figure}
\centerline{\includegraphics[width=2.8in]{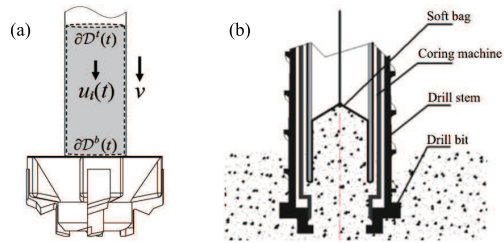}}
\caption{Schematic of the internal channel. The dashed line represents the
boundary $\partial\mathcal{D}$ of the domain $\mathcal{D}$ filled by the particles in the internal channel. (b) defines various components of the coring tube, including the soft bag used to collect the soil sample.}
\label{fig:sketch}
\end{figure}

The sampling process in the internal channel concerns a domain $\mathcal{D}$ with its two moving boundaries: the bottom $\partial \mathcal{D}^b$ and the top $\partial \mathcal{D}^t$ surfaces (see figure~\ref{fig:sketch}). $\partial \mathcal{D}^b$  moves downward with penetration velocity $v$, whereas $\partial \mathcal{D}^t$ takes the same velocity $u_i$ as the granular sample, assuming collective motion of all grains in the internal channel. As sketched in Fig.\,\ref{fig:sketch}(b), one end of the soft bag is held firmly via an attached string. During the drilling process, the sample is being collected in the soft bag as the coring tube penetrates deeper into the lunar simulant. Since the normal stress between the granular material and the bag is relatively small in comparison to that in the outer channel, we neglect the frictional force between the granular sample and the inner tube. Granular flow in the inner channel can be considered as a one-dimensional flow. More details on the functionality of the soft bag can be found in \cite{Tang2017}.

The mass sampled in the internal channel $m_i$ is also time-dependent, and its rate of change is governed by
\begin{equation}\label{eq:inmassrate}
  \frac{\ud m_i}{\ud t}=\frac{\ud}{\ud t}\int_{\mathcal{D}}\rho \ud V = \oint_{\partial \mathcal{D}}\rho v_n \ud S=\rho (v-u_i)S_i
\end{equation}
where  $v_n$ is the velocity normal to the surface  $\partial \mathcal{D}$ of the domain $\mathcal{D}$, and $S_i=\pi r_i^2$ denotes the cross-sectional area of the internal channel.

As the rate of momentum change for fluid in volume $\mathcal{D}$ must be balanced by body force and surface pressure. The integral momentum balance reads
\begin{equation}\label{eq:inbalance}
  \frac{\ud }{\ud t}\int_{\mathcal{D}}\rho {u_i} \ud V=\int_{\mathcal{D}} \rho {g} \ud V + \oint_{\partial \mathcal{D}} {p_i} \ud S
\end{equation}
where ${p_i}$ is the normal pressure on the surface $\partial \mathcal{D}$.

The left side of Eq.~\ref{eq:inbalance} satisfies
\begin{equation}\label{eq:lcs-1}
\begin{split}
  \frac{\ud }{\ud t}\int_{\mathcal{D}}\rho {u_i} \ud V &=\int_{\mathcal{D}} \rho \frac{\partial{u_i}}{\partial{t}}\ud V+ \oint_{\partial \mathcal{D}} \rho {u_i} v_n \ud S\\
       &= m_i \frac{\ud u_i}{\ud t} + \rho S_i u_i(v-u_i).\\
\end{split}
\end{equation}

\noindent $P$ denotes the normal pressure exerted on boundary $\partial \mathcal{D}^b$. Note that the top boundary $\partial \mathcal{D}^t$ is a free surface. The second term of the right side of Eq.~\ref{eq:inbalance} can be written as
\begin{equation}\label{eq:lcs-2}
  \oint_{\partial \mathcal{D}} {p_i} dS = -PS_i.
\end{equation}

\noindent Thus, Eq.(\ref{eq:inbalance}) can be expressed as
\begin{equation}\label{eq:inacc}
  m_i \frac{\ud u_i}{\ud t} =m_i g-PS_i-\rho S_i u_i(v-u_i).
\end{equation}

In summary, granular flow in the internal channel is governed by the continuum equation Eq.~(\ref{eq:inmassrate}) and the momentum balance equation  Eq.~(\ref{eq:inbalance}) with three time-dependent variables: $m_i$, $u_i$ and $P$.

\subsection{Flow dynamics in external channel}
Figure~\ref{fig:outvela} shows the central layer of the equivalent chute flow along the streamwise direction. Note that the helical motion of granular sample in the external channel is similar to a granular chute flow with inclination angle $\beta$, considering the auger flight being unwrapped. In the flowing layer,  we establish a coordinate system  with unit vectors $\hat{\bm{s}}$ and $\hat{\bm{\xi}}$ representing the streamwise and normal directions, respectively.  We assume that the granular sample fills up the helix groove along $\hat{\bm{\xi}}$ direction, but the flow thickness $\eta$ is smaller than the groove depth $a$ (see figure~\ref{fig:Config}(b)) to account for the loss of materials due to mass exchange between the external channel and surroundings. Therefore, the central layer of the flow has a radial distance $\overline{r}=r_i+\eta/2$, and an inclination angle $\beta=b/(2\pi\overline{r})=b/[\pi(2r_i+\eta)]$. Because the thickness of the coring tube is relatively small in comparison to $r_{\rm i}$ or $\eta$, it is neglected in the current investigation. The width of the flowing layer is then computed as $\xi= b\cos\beta$, and the cross-sectional area of the external channel is given by $S_o=\xi\eta=\eta b\cos\beta$.

The discharging process of granular flow in the external channel is related to a time-dependent domain $\mathcal{R}$ with two boundaries: the fixed boundary $\partial \mathcal{R}^u$ on the soil surface, and a moving boundary $\partial \mathcal{R}^d$ on the bottom surface of the drilling hole (see figure~\ref{fig:outvela}). Note that the boundary $\partial \mathcal{R}^d$  moves with the drilling velocity $v$ and the area of $\partial \mathcal{R}^d$ is $S_o/\sin\beta$ according to the geometrical relationship shown in figure~\ref{fig:outvela}. The mass of grains flowing into the external channel is $m_o$, and its rate of change is governed by
\begin{equation}\label{eq:outmassrate}
  \frac{\ud m_o}{\ud t}=\frac{\ud }{\ud t}\int_{\mathcal{R}}\rho \ud V = \oint_{\partial \mathcal{R}}\rho v_n \ud S =\rho v \frac{S_o}{\sin\beta},
\end{equation}
which is a constant under the conditions of constant penetration velocity $v$, cross-sectional area $S_o$, and bulk density $\rho$. Thus, we have $m_o = \rho S_o v t /\sin\beta$.

Similar to the case of internal channel, we assume homogeneous flow velocity $u_s$ along the streamwise direction in the external channel. Momentum balance can then be expressed as
\begin{equation}\label{eq:outbalanceproj}
\frac{\ud }{\ud t}\int_{\mathcal{R}}\rho u_s \ud V =-\int_{\mathcal{R}} \rho g \sin\beta \ud V  + \oint_{\partial \mathcal{R}} t_o(s) \ud S
\end{equation}
where $t_o(s)$ is the surface stress exerted on the boundary ${\partial \mathcal{R}}$. Note that the left side of Eq.~\ref{eq:outbalanceproj} can be written as
\begin{equation}\label{eq:outbalanceleft}
  \begin{split}
\frac{\ud }{\ud t}\int_{\mathcal{R}}\rho u_s \ud V &=\int_{\mathcal{R}} \rho \frac{\partial{u_s}}{\partial t}\ud V  + \oint_{\partial \mathcal{R}} \rho u_s v_n \ud S \\
&=m_o \frac{\ud  u_s}{\ud t}+ \rho v u_s \frac{S_o }{\sin\beta}
  \end{split}
\end{equation}
with surface normal pressure $P$ exerted on boundary $\partial \mathcal{R}^d$. Hence,  Eq.~\ref{eq:outbalanceproj}  can be rewritten as
\begin{equation}\label{eq:outacc}
m_o \frac{\ud u_s}{\ud t}= -m_o g\sin{\beta}  -  \frac{\rho S_o v u_s}{\sin\beta} + PS_o  +Fr,
\end{equation}
where $Fr$ is the frictional force exerted on the lateral surfaces of the flowing layer. It plays an important role in determining the granular flow in the external channel, which is discussed in details as follows.

\subsection{Frictional force on the flow in external channel}
The flowing layer enclosed in the domain $\mathcal{R}$ of the external channel is subjected to gravity, centrifugal force and surface stresses. We need to consider internal stress and lateral friction in describing the flowing layer.

\begin{figure}
  \centerline{\includegraphics[width=5.0in]{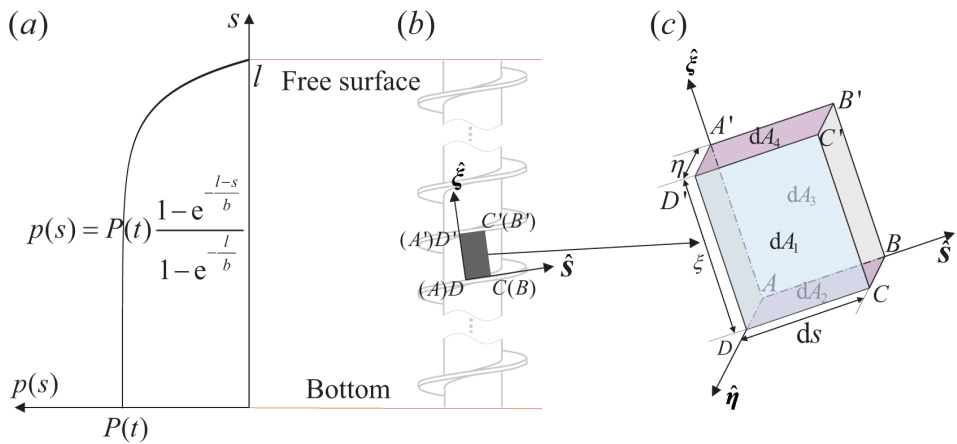}}
  \caption{(\textit{a}) The profile of the  pressure distribution along the external channel.
    (\textit{b}) An infinitesimal element of the flow on the auger flight. (\textit{c}) The
    surfaces of the infinitesimal element plotted in a local coordinate frame $A-\hat{\bm{s}}\hat{\bm{\xi}}\hat{\bm{\eta}}$,
    where $\hat{\bm{\eta}}=\hat{\bm{s}}\times \hat{\bm{\xi}}$.}\label{fig:stress}
\end{figure}

To analyze the frictional force on the lateral surfaces of the flowing layer, we select an infinitesimal hexahedron element $(\xi\times \eta \times \ud s)$, as illustrated in figure~\ref{fig:stress}(b) and (c). There are four lateral surfaces designated by $\ud A_j $ ($j$=1,2,3,4).  $\ud A_1 $ and $\ud A_3 $ are the lateral surfaces in touch with the surrounding static granular materials and the groove bottom, respectively. Their areas are computed as $\ud A_1 =\ud A_3 =\xi \times \ud s$.  $\ud A_2$ and $\ud A_4$ are the lateral surfaces contacting  the top and bottom surfaces of the auger flight, respectively, and $\ud A_2 =\ud A_4 =\eta \times \ud s$. The magnitudes of shear and normal stresses on each lateral surface $A_j$, $j=1, 2, 3, 4$,  are denoted as  ${\tau_j}(s)$ and ${\sigma_j(s)}$, respectively.

As the sample in the external channel moves upwards with a fixed elevation angle $\beta$, it is reminiscent to a chut flow with additional centrifugal force in the radial direction. In this configuration, assuming the shear stress on surface $A_1$ satisfies the $\mu{}-{}I$ rheology introduced in \citet{jop2006constitutive,midi2004dense}, we have
\begin{equation}\label{eq:granularfriction}
  {\tau_1} (s) = \mu (I){\sigma_1}(s),
\end{equation}
where $\mu (I)=\mu_s+(\mu_2-\mu_s)/(I_0/I+1)$ with $I_0$, $\mu_s$ and $\mu_2$ model parameters. Inertial number $I$ is estimated with $I=\dot{\gamma_{\rm s}}d/(\sigma_1(s)/\rho)^{1/2}$, where $d$ is the average grain diameter, and $\dot{\gamma_{\rm s}}=\sqrt{u^2_{s}+u^2_{\xi}}/(N d)$ denotes the shear rate for a shear band with thickness $N d$. The granular friction coefficient $\mu (I)$ starts with a critical value $\mu_s$ at zero shear rate, increases with inertial number $I$ and eventually converges to a finite value $\mu_2$. Note that the variation of $\mu$ is not significant in the steady state because of the stable inertial number for the parameter range explored here, thus one may also assume a constant $\mu$ as a first approximation. Nevertheless, the fluctuation of $\mu$ with $I$ can be significant in the initial transient state. Here, the velocity of the flowing layer is $\bm{u} = u_s \hat{\bm{s}} +u_\xi \hat{\bm{\xi}}$. 

As the direction of the shear stress ${\tau_1} (s)$ is opposite to that of $\bm{u}$, we can decompose the shear stress on surface $A_1$ along the streamwise and normal directions as follows:
\begin{equation}\label{eq:tau1}
\bm{\tau_1} (s) = {\tau_1^s} (s)\cdot \hat{\bm{s}} + {\tau_1^\xi} (s)\cdot \hat{\bm{\xi}} = -\left(\cos\theta\cdot\hat{\bm{s}} + \sin\theta \cdot \hat{\bm{\xi}}\right) \mu (I){\sigma_1}(s)
\end{equation}
with $\theta=\tan^{-1}(u_\xi/u_s)$.

The normal stress $\sigma_1(s)$ on lateral surface $A_1$ arises from the hydrostatic pressure $p(s)$ and the pressure $p_c$ arising from the centrifugal force $F_c$:
\begin{equation}\label{eq:lateralpressure}
\begin{split}
  \sigma_1(s)=p(s)+p_c.
\end{split}
\end{equation}

Note that the flow along the normal direction is constrained by the auger flight, which is subject to a compound motion of penetration and rotation. This means that, as the flow is considered to be incompressible, the component $u_{\xi}$ equals the normal component of the auger's velocity $\bm{v}\cdot\hat{\bm{\xi}}$. Hence, we have
\begin{equation}\label{eq:mu_n}
u_{\xi}=\omega \overline{r} \sin{\beta}-v\cos{\beta},
\end{equation}
and $u_\xi$ should be a constant when the auger's motion is given.
The circumferential speed around the rotation axis of the flowing layer is expressed as  $u_{h} = u_\xi \sin \beta - u_s \cos\beta$. Therefore, the centrifugal force per mass $\ud m$ reads
\begin{equation}\label{eq:centrifugalforce}
  F_c=\ud m \frac{u_{h}^2}{\overline{r}} = \ud m \frac{(u_\xi\sin\beta-u_s\cos\beta)^{2} }{\overline{r}},
\end{equation}
where $\ud m  = \rho \xi\eta \ud s $ and further $p_c=F_c/\ud A_1$.

Suppose that the region around the drill bit has isobaric pressure. The normal pressure exerted on the boundary $\partial \mathcal{R}^b$ has the same surface pressure $P$ as that on boundary  $\partial \mathcal{D}^b$. Noting that boundary $\partial \mathcal{R}^u$ corresponds to a free surface of the soil, its surface pressure equals zero. Considering the Janssen effect~\citep{Duran2000}, we assume that the surface pressure of the flowing layer along the streamwise direction is exponentially distributed with $s$, namely, the hydrostatic pressure at position $s$ is expressed as
\begin{equation}\label{eq:pre_dis}
p(s)=P\frac{1-\ue^{-\frac{l-s}{b}}}{1-\ue^{-\frac{l}{b}}},
\end{equation}
where $l$ is the total length of the external channel immersed in the soil at time $t$. Figure~\ref{fig:stress}(a) shows the  distribution of hydrostatic pressure $p(s)$, in which  $p(s=0)=P$ and $p(s=l)=0$, at the bottom of the drill stem and at the free surface of the soil, respectively.

According to the equilibrium condition on surface $A_2$, the normal stress ${\sigma_2}(s)$ can be estimated with
\begin{equation}\label{eq:sigma2}
    {\sigma_2}(s)= p(s)+\rho g b \cos^2{\beta} + {\tau_1^\xi} (s)\frac{\xi}{\eta}.
\end{equation}
It is composed of hydrostatic-like pressure $p(s)$, gravity and the normal component  ${\tau_1^\xi} (s)$ of the friction force on surface $A_1$.

The normal stresses on surfaces $A_3$ and $A_4$ are induced by the hydrostatic-like pressure $p(s)$, i.e., ${\sigma_3}={\sigma_4}=p(s)$. Note that the flowing layer takes a relative motion along the streamwise direction $\hat{\bm{s}}$ with respect to the lateral surfaces $A_2$, $A_3$ and $A_4$. Assuming that the shear stresses on these three surfaces satisfy the Coulomb friction law, we have
\begin{equation}\label{eq:tau2}
    {\tau_j}(s)=-\mu_0 {\sigma_j}(s), \;\;\; j= 2, 3, 4,
\end{equation}
where $\mu_0$ is the friction coefficient between granular materials and the drill stem surface.

Finally, we integrate the four shear stresses along the streamwise direction $\hat{\bm{s}}$ to obtain the frictional forces exerted on the flowing layer.
\begin{equation}\label{eq:friction}
Fr = \int_0^{l} (\tau_1^s (s) +  {\tau_3}(s))  \xi \ud s + \int_0^{l} (\tau_2(s) + \tau_4(s)) \eta \ud s.
\end{equation}

It shows that frictional force $F_r$ primarily arises for hydrostatic pressure, geometry, and frictional coefficient $\mu$ that depends on inertial number.

\subsection{Coupling between internal and external channels}
The granular material generated by the drill bit either flows into the internal channel or is conveyed by the external channel. The mass increase rate is given by $\ud m_b/\ud t=S_bv$ with $S_b$ the cross-section area at the bottom of the drill, where granular flow is generated by the bit. Here, $S_b$ is estimated with a summation of the bottom areas of both internal and external channels:

\begin{equation}\label{eq:areas}
     S_b  =  S_i + \frac{S_o}{\sin\beta}.
\end{equation}

The increased mass of granular materials in the internal and external channels are given by Eqs. (\ref{eq:inmassrate}) and (\ref{eq:outmassrate}), respectively. Meanwhile, part of the granular materials are removed to the soil surface through the external channel.  The velocity of the removed soil on the top boundary $\partial \mathcal{R}^u$ of  the flowing layer can be computed as  $u_{\text{up}} =u_{\xi}\cos\beta+u_s\sin\beta$.  Thus, the rate of mass removal  by the external channel is given by $\ud m_{\text{rem}}/\ud t = \rho S_o u_{\text{up}}/\sin\beta$. According to mass conservation in both channels, we have
\begin{equation}\label{eq:massconservation}
     S_b v = (v-u_i)S_i +  \frac{v +  u_{\text{up}}}{\sin\beta}{S_o}.
\end{equation}
Together with the definition of $u_{\xi}$, the second term can be written as known variables. Subsequently, the above equation for the mass conservation in the two channels can be transformed into the following form:
\begin{equation}\label{eq:volumeconservation}
  S_b v=  S_o (v\sin\beta + \omega  \overline{r} \cos\beta + u_s) + S_i (v-u_i),
\end{equation}
Eq.\eqref{eq:volumeconservation} can also be considered as a kinematic constraint that couples the flow in the internal and external channels.

\subsection{Summary of the continuum model}
The above analysis yields five governing equations with five time-dependent variables $m_i$, $m_o$, $u_i$, $u_s$, and $P$. More specifically, there are two mass balance equations (Eqs.~(\ref{eq:inmassrate}) and (\ref{eq:outmassrate})), two momentum balance equations ( Eqs.~(\ref{eq:inacc}) and (\ref{eq:outacc})), and coupling equation shown in Eq.~(\ref{eq:massconservation}). Note that $F_r$ can be represented as a function of $P$. Given a certain model parameters and initial conditions, the granular flow in both channels can be described numerically.

Differentiating Eq.~(\ref{eq:volumeconservation}) leads to
\begin{equation}\label{eq:us_ui}
  \frac{\ud u_s}{\ud u_i} = \frac{S_i}{S_o}.
\end{equation}

Combining the above equation with Eqs. (\ref{eq:inacc}) and (\ref{eq:outacc}) allows us to write the analytical expression of the hydrostatic-like pressure $P$:
\begin{equation}\label{eq:pressure}
  P=\frac{m_i S_o C_1+m_o S_i C_2}{m_i S_o^2+m_o S_i^2},
\end{equation}
where $C_1 \equiv m_o g\sin{\beta}  +  \rho S_0 v u_s/\sin\beta - Fr$ and $C_2 \equiv m_i g-\rho S_i u_i(v-u_i)$.

Suppose $m_i(t_0) =0$ and $m_o(t_0) = 0$  at time $t_0=0$.  The two continuum equations in  both channels can be  written in the following integral forms:
\begin{equation}\label{eq:massrate}
\left \{
  \begin{aligned}
 &m_i = \rho S_i \left(v t -\int_0^t u_i\ud t\right),\\
 &m_o  =  \frac{\rho S_ov}{\sin\beta}t.
  \end{aligned}
  \right.
\end{equation}

The dynamics of the granular flow in  both channels is governed by Eqs.~(\ref{eq:inacc}) and (\ref{eq:outacc}), which are reorganized here for clarity.
\begin{equation}\label{eq:unidyna}
\left \{
  \begin{aligned}
 &\frac{\ud u_i}{\ud t} = g-\frac{P}{ m_i}S_i-\rho S_i \frac{u_i(v-u_i)}{m_i},\\
 &\frac{\ud u_s}{\ud t}= - g\sin{\beta}  -  \frac{u_s}{t} + \frac{P\sin\beta}{\rho v t}+ \frac{Fr\sin\beta}{\rho S_o vt}.
  \end{aligned}
  \right.
\end{equation}

Note that the two equations are subject to the kinematic constraint given by Eq.~(\ref{eq:volumeconservation}). Therefore, the initial values of $u_i(t_0)$ and $u_s(t_0)$ cannot be specified arbitrarily, but they should satisfy the kinematic condition. For instance, if we set $u_i(t_0)=0$, then the value of $u_s(t_0)$  should be computed by Eq.~(\ref{eq:volumeconservation}) such that the condition of mass conservation can be satisfied at time $t_0$. It clearly suggests that gravity plays an important role in $u_i$, and consequently on the sampling efficiency. Based on the initial values of  $u_i(t_0)$ and $u_s(t_0)$, together with Eqs.~(\ref{eq:pressure}), (\ref{eq:massrate}) and (\ref{eq:unidyna}), we can numerically obtain the solutions of $u_i$, $u_s$, $m_i$, $m_o$, and $P$.

\section{Validation of the continuum model}\label{sec:validation}
\begin{table}
\begin{center}
\def~{\hphantom{0}}
\vspace{-9 pt}
\caption{Parameters used in the theoretical model are selected to match experimental conditions, including $\beta (^{\circ}) =\frac{180 b}{\pi^2(2r_i+\eta)}$, $r_i$ = 1.55(cm), b=1.2(cm), $\xi=b\cos\beta$, $S_i=\pi r_i^2$, $S_o=\xi\eta$.}\label{tab:exp}
\begin{tabular}{cccccccccc}
  \toprule
  Soil type & $\rho$(g/cm$^3$) & $\mu_s$ &$\mu_0$  & $\mu_2$ &$\eta$(cm)  & $\xi$ (cm) &   $\beta$ ($^\circ$)\\
  \midrule
  Soil-I     &  2.13     & 0.72      & 0.57  & 0.8   &0.025 &1.19  &6.97\\
  Soil-II    &  1.85     & 0.60      & 0.46 & 0.7   &0.105 &1.19  &6.81\\
  \bottomrule
  \vspace{1 pt}
\end{tabular}
\end{center}
\end{table}

In this section, we verify the continuum model through a comparison with experimental results shown in \S~\ref{sec:experiment}. Parameters used to numerically solve the governing equations described above are listed in table~\ref{tab:exp}. They are chosen based on experimental conditions as discussed below.

The bulk density $\rho$ is chosen to match experimentally measured ratio of sample weight over volume occupied. The internal friction coefficient $\mu_s = \tan \phi$ is determined from the angle of repose $\phi$ of the granular sample obtained after the drilling process to reflect properties of granular sample in the fluidized state. We assume that the granular flow satisfies Mohr-Coulomb yield criterion~\citep{kang2018archimedes,feng2019support}. The frictional coefficient between the granular sample and the surfaces of the auger groove, $\mu_o$ is measured using standard slip testing method (GB/T 22895-2008). In order to estimate $\mu(I)$ with Eq.\eqref{eq:granularfriction}, we need three parameters: The asymptotic frictional coefficient $\mu_2$, the thickness of shear band $Nd$ and the material-dependent constant $I_0$. We set $N=10$ and use $I_0= \frac{0.206}{\sqrt{\psi \cos\beta}}$ with $0.206$ a material dependent constant and $\psi$ packing fraction of the granular sample to estimate the inertial number, following previous investigations~\citep{jop2005crucial,Forterre2003}. For Soil-I and Soil-II, we have $I_0= 0.42, 0.58$, respectively.

Parameters $\xi$ and $\eta$ correspond to the width and thickness of the flowing layer in the inclined external channel. Note that granular flow in the external channel is confined in a shallow auger groove and presumably limited to few grain diameters~\citep{jop2005crucial,midi2004dense,wang2019velocity}. As the flow takes place at the interface between sample being displaced by the drill and surrounding soil underground, it is challenging to measure it experimentally. Instead, we obtain $\eta$ through fits to the experimental data. As shown in table~\ref{tab:exp}, fitting shows that $\eta$ is on the order of one or two particle diameters, in agreement with the above analysis.
\begin{figure}
\centering
\begin{tabular}{cc}
\subfigure{\includegraphics[width=0.48\linewidth]{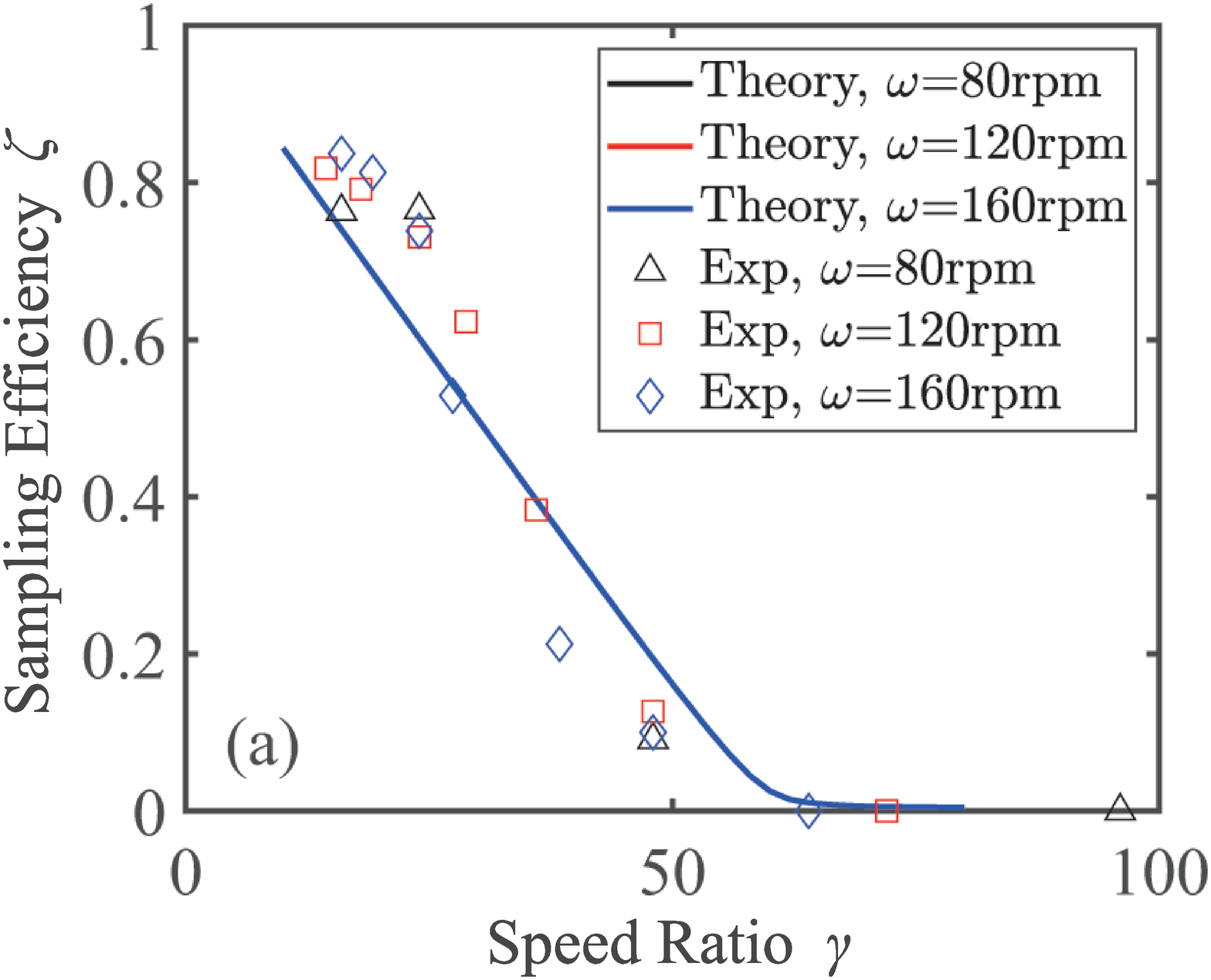}\label{fig:comparison1}}&
\subfigure{\includegraphics[width=0.48\linewidth]{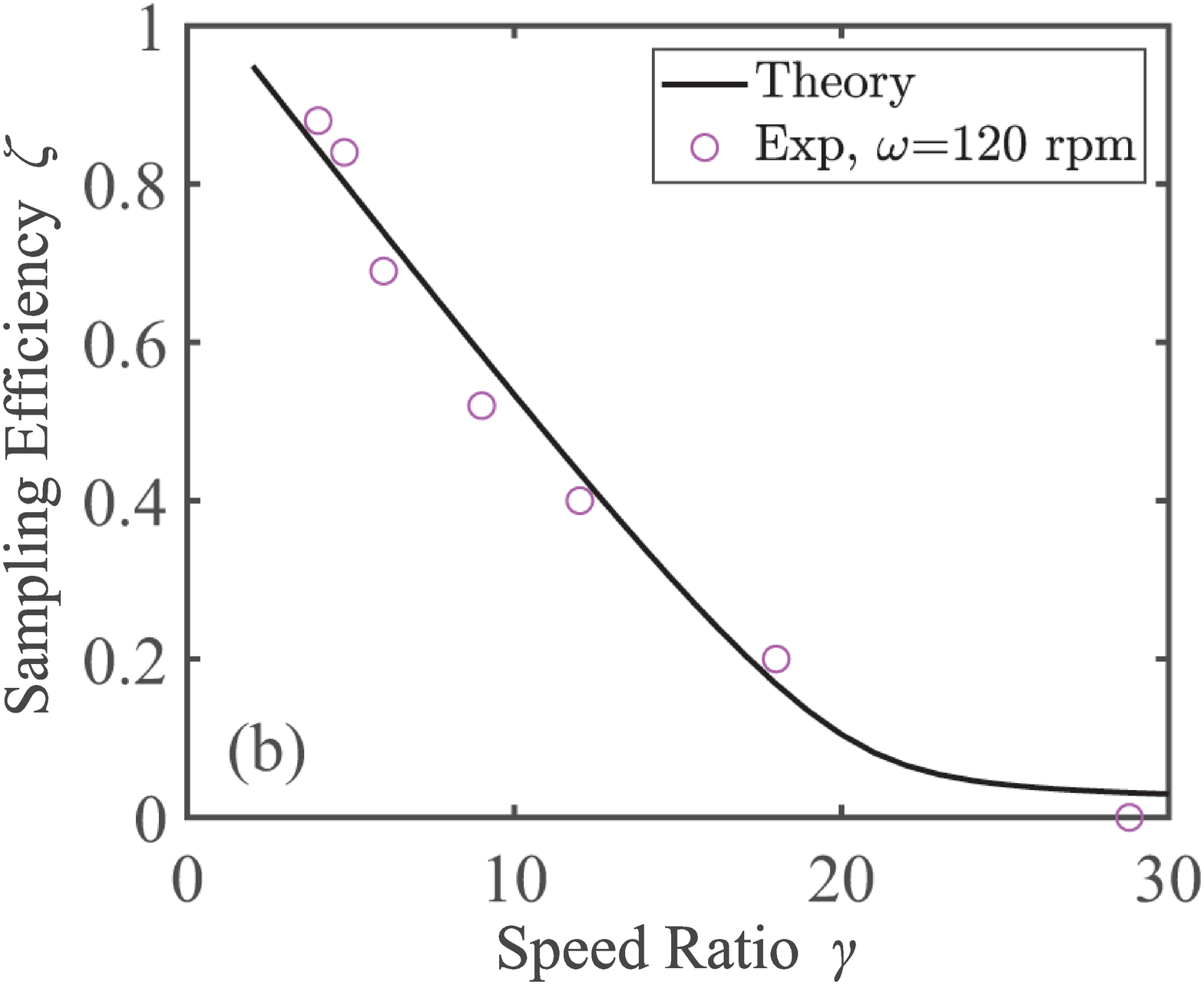}\label{fig:comparison2}}
\end{tabular}
\caption{Comparison between the numerical and experimental results for the relationships
between sampling efficiency $\zeta$ and speed ratio $\gamma$ when the auger drills in Soil-I (\textit{a}),
and Soil-II (\textit{b}).}
\label{fig:expcomparison}
\end{figure}

Based on the parameters listed in Table~\ref{tab:exp}, we  employ Eqs.\eqref{eq:pressure}, \eqref{eq:massrate} and \eqref{eq:unidyna} to simulate the process of auger penetrating into the two types of soils under different driving conditions. All the simulations are terminated at the time when the penetration depth $H=1.0$\,m is achieved. Numerically, we obtain the mass $m_i$ of sampled granular materials at the end of each drill process. Subsequently, Eq.~\eqref{eq:samplingrate1} is used to estimate $\zeta$. How the sampling efficiency $\zeta$ varies as a function of $\gamma$ is investigated by variations of the penetration speed $v$ and rotational speed $\omega$ following experimental conditions. As shown in figure~\ref{fig:comparison1}, the $\zeta-\gamma$ curves for Soil-I under different rotational speeds overlap with each other well, clearly demonstrating that the sampling efficiency $\zeta$ depends on the speed ratio $\gamma$ instead of specific rotational speeds. Figure~\ref{fig:expcomparison} also shows good agreements between numerical and experimental results for both types of soils within the parameter range explored here. In the future, down-scaled experiments capable of exploring a wider range of $\gamma$ with finer steps are needed to further validate the model. Note that $\zeta$ decays asymptotically to zero as $\gamma$ grows to infinity for both types of lunar simulants. This extreme case corresponds to infinitely large $\omega$ at given $b$ and $v$, consequently the centrifugal force induced by the rotation of the drill effectively enhances the normal forces applied on the walls of the external channel. Thus, the effective friction is large to prohibit the flow of granular sample, and the percentage of sample being filled in the inner tube decreases asymptotically to 0. Nevertheless, it cannot reach 0 for the limited parameter range explored here, and the pressure on the bottom of the drill $P$ is always positive (see Fig. \,\ref{fig:liu-2}(b)).”
\begin{figure}
\centering
\begin{tabular}{cc}
\subfigure{\includegraphics[width=0.48\linewidth]{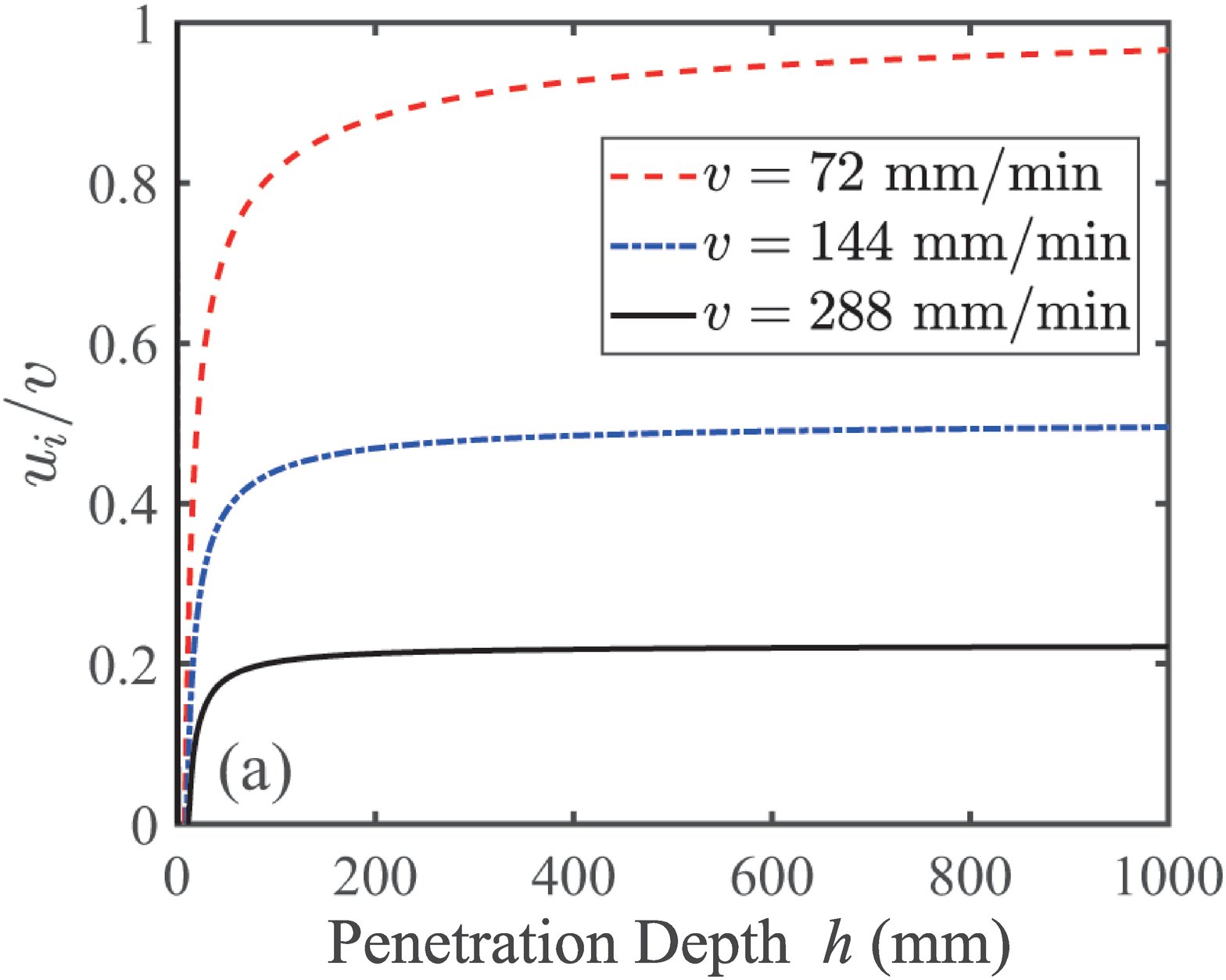}\label{fig:invel}} &
\subfigure{\includegraphics[width=0.48\linewidth]{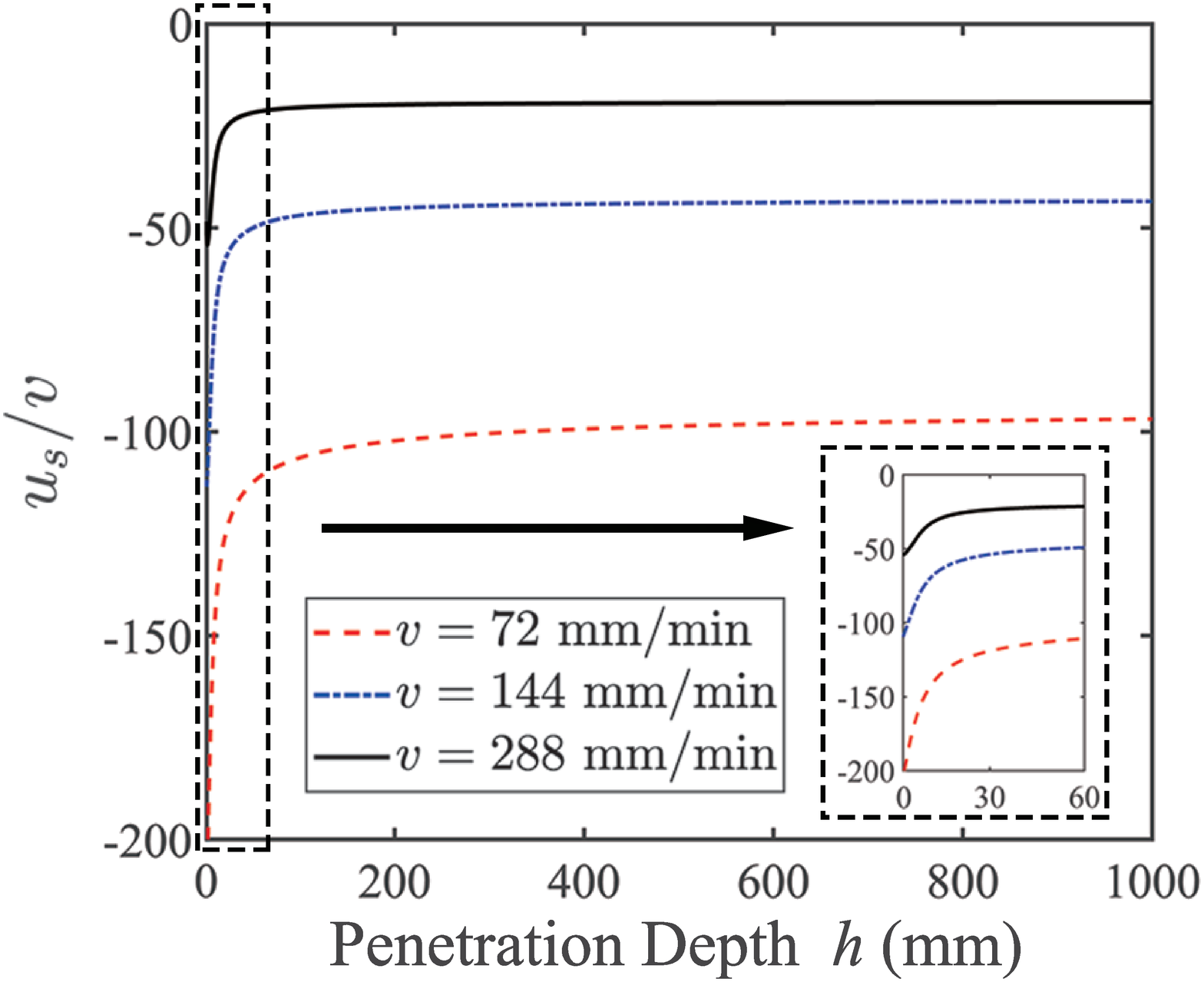}\label{fig:outvel}}
\end{tabular}
\caption{Dimensionless flow speed $u_i/v$ in the internal channel (\textit{a}), as well as the corresponding dimensionless velocity $u_s/v$ in the external channel (\textit{b}), as a function of the penetration depth $h = vt$. Here, $u_s/v$ is negative because the granular surface of the fluidized sample moves upwards along the external channel, i.e., in a different direction as the drill. Simulations are performed for drilling into Soil-II with fixed $\omega=120$rpm and three penetration speeds $v=72, 144, 288$~mm/min. Inset of (b) shows a close-view of the velocity change at the very beginning of the penetration process.}
\label{fig:IOvel}
\end{figure}

Based on the above comparison to experimental data, we further analyze granular flow dynamics in both internal and external channels of the auger drill. As shown in figure~\ref{fig:IOvel}, numerical results for the case of drilling in Soil-II with a fixed $\omega$ and three different penetration speeds $v$ clearly suggest that both flow speeds in internal and external channels converge to a constant value as the penetration depth $h$ increases (see figure~\ref{fig:IOvel}), suggesting the existence of a steady state with constant granular flow speeds in the internal and external channels as time evolves. Qualitatively, figure~\ref{fig:IOvel} also reveals that larger penetration speeds lead to quicker converging into the steady state.

\begin{figure}
\centering
\begin{tabular}{cc}
\subfigure{\includegraphics[width=0.48\linewidth]{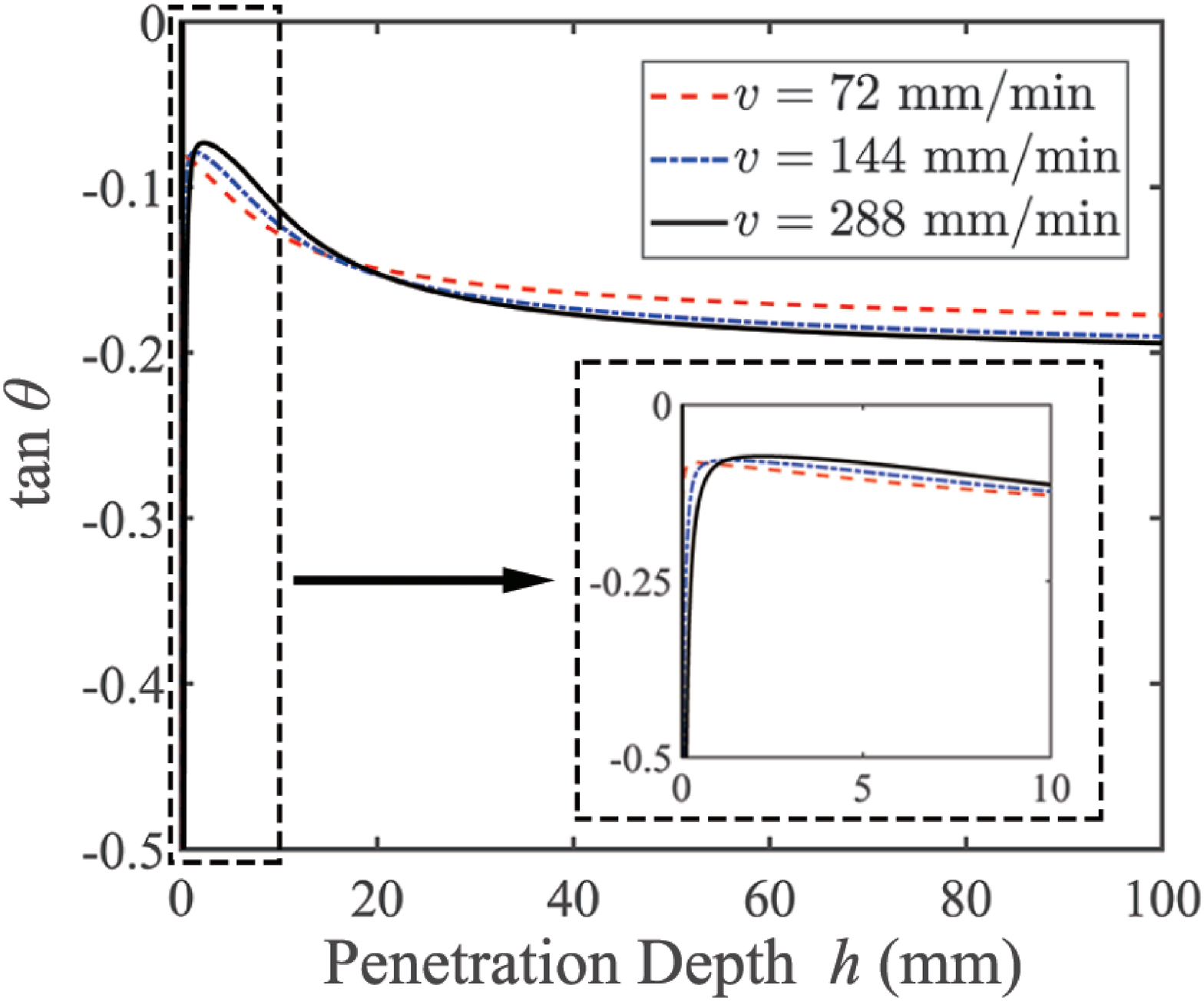}\label{fig:tantheta}} &
\subfigure{\includegraphics[width=0.48\linewidth]{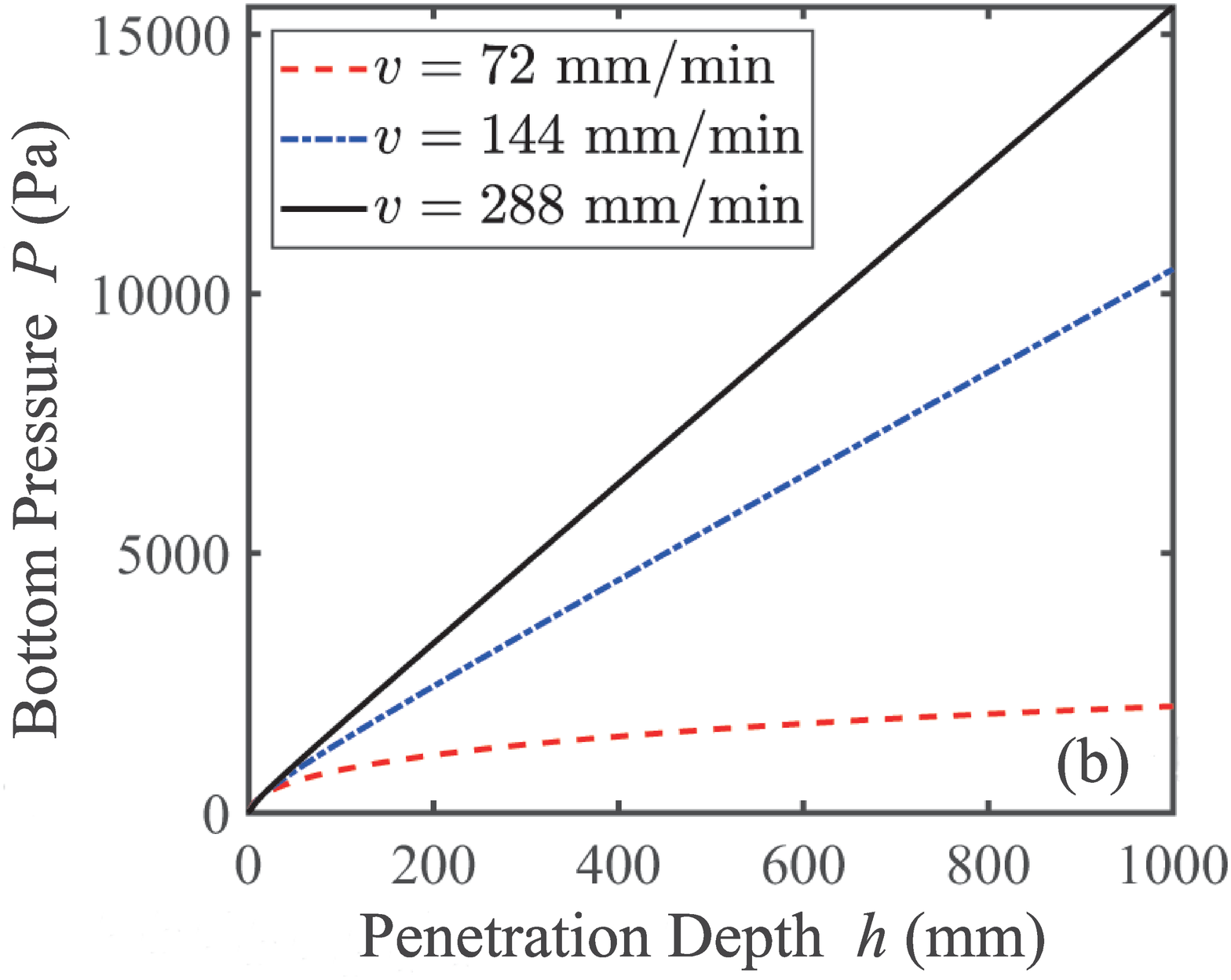}\label{fig:bottomepressure}}
\end{tabular}
\caption{Direction of external flow velocity $\tan\theta=u_\xi/u_s$ (a), and
pressure $P$ (b) as a function of penetration depth $h$ for the corresponding conditions shown in figure~\ref{fig:IOvel}. Inset of (a) is a close-view of the angle change at small $h$.}
\label{fig:liu-2}
\end{figure}

In addition to granular flow along the streamwise direction $u_s$, Eq.\eqref{eq:mu_n} indicates the other flow component $u_\xi$ generated from the helical motion of the drilling tool. As shown in figure~\ref{fig:tantheta}, the direction of granular flow $\tan\theta$ evolves quickly into a constant value as $h$ grows, despite of the peaks emerging at small $h$ (see the inset). Within the parameter range explored, $\tan\theta$ is independent of the driving conditions and becomes steady at the very initial stage of penetration ($h\le 50$\,mm).  Moreover, the corresponding evolution of the bottom pressure $P$ during the penetration process under different driving conditions is plotted in figure~\ref{fig:bottomepressure}. It shows that $P$ grows linearly with the penetration depth in the steady state, in which both $u_s$ and $\theta$ become stable. In the steady state, the growth rate increases at larger $v$. 

\begin{figure}
\centering
\begin{tabular}{cc}
\subfigure{\includegraphics[width=0.48\linewidth] {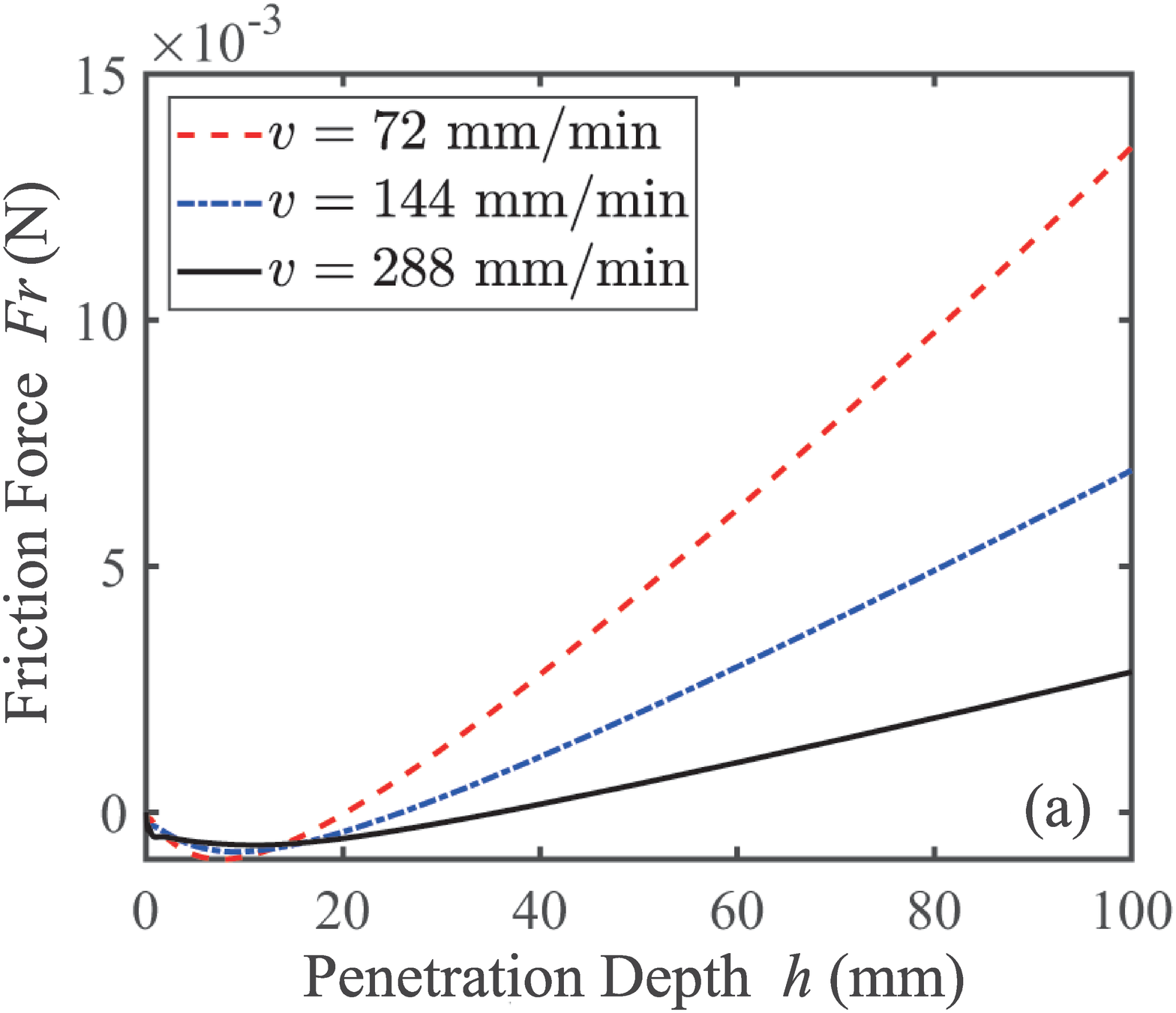}\label{fig:Fr}}&
\subfigure{\includegraphics[width=0.48\linewidth]{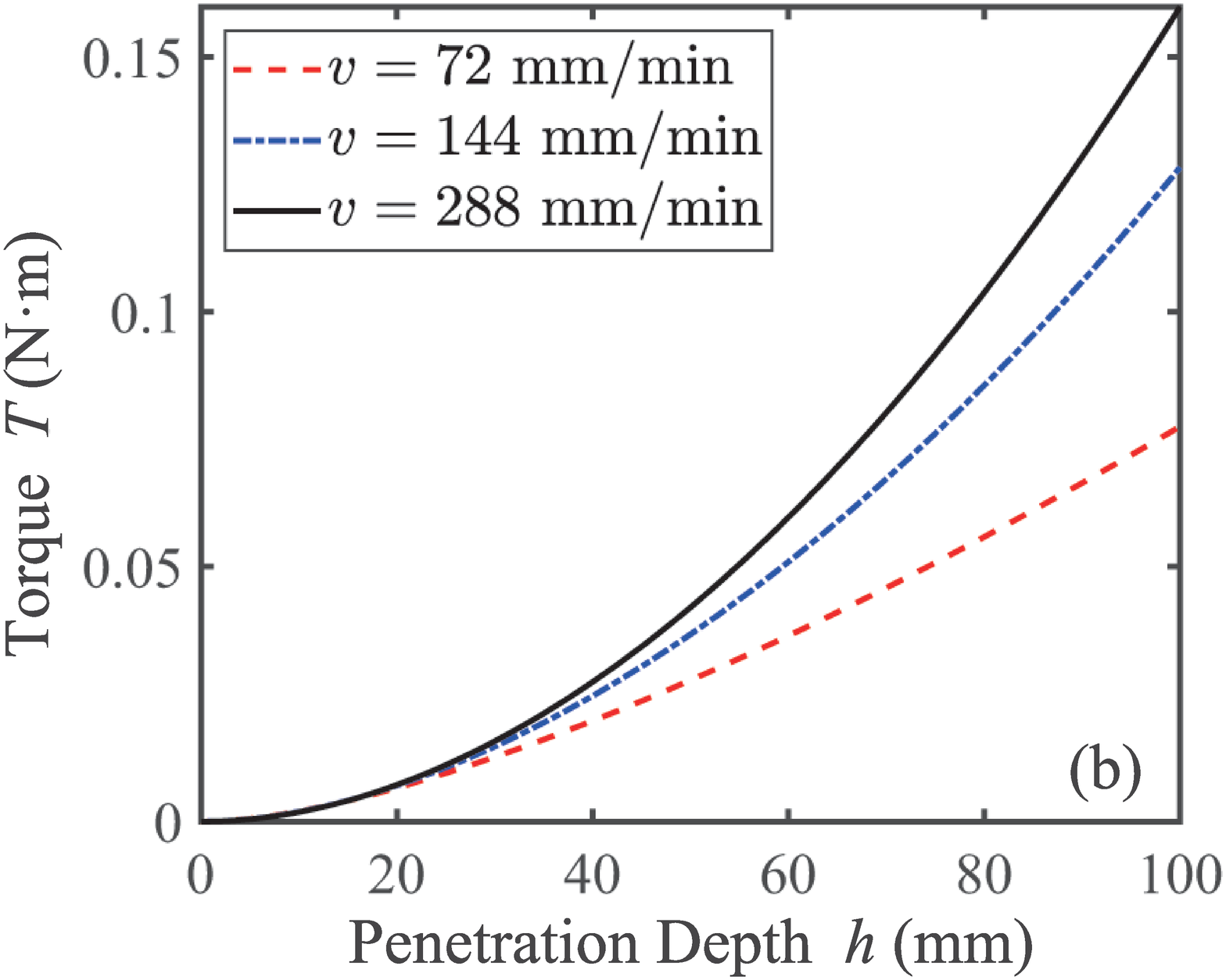}\label{fig:tor}}
\end{tabular}
\caption{(a) the friction force $Fr$ exerted on the flowing layer, and (b) the resistant
torque arising from the friction between the flowing layer and the surrounding static granular
 material, vary with the penetration depth $h$. The same  driving conditions as   shown
 in the caption of figure~\ref{fig:IOvel} are employed in the simulations for drilling into  Soil-II.}
\label{fig:liu-3}
\end{figure}

According to Eq.\eqref{eq:friction}, we analyze the friction force $Fr$ exerted on the flowing layer in the external channel as a function of with drilling depth $h$ under three different penetration speeds. Figure~\ref{fig:Fr} shows that the frictional force decays initially to a negative value as $h$ increases, owing to the non-monotonic behavior of $u_s$ in the initial stage of penetration. As the system approaches the steady state with constant $u_s$, $Fr$ grows approximately linearly with penetration depth, in relation to the growth of $P$ with $h$. Notably, frictional force decreases with increasing penetration speed. From previous analysis, we know that the granular flow direction in the external channel is opposite to the vector $\hat{\bm{s}}$ (i.e., negative $u_s$) and tends to be constant as the $h$ increases. From equation~\ref{eq:outacc}, we know that the right hand side approaches $0$ in the steady state. Consequently, the increase of both the second ($ -\rho S_o v u_s/\sin\beta$) and third terms $P S_0$ with $v$ in the steady state leads to the decrease of $Fr$ as penetration speed $v$ grows.

As granular sample is being conveyed upwards through the external channel, there exists a torque $T$ arising from the frictional force between the sample and surrounding granular materials:
\begin{equation}\label{eq:force}
  \begin{split}
       T=\int_0^{l} r_o \hat{\bm{\eta}} \times \bm{\tau_1} b \cos\beta \ud s \cdot \hat{\bm{k}}
  \end{split}
\end{equation}
where $l$ is the length of granular layer in the external channel and $\hat{\bm{k}}$ is the axial unit vector of the drill stem. As shown in figure~\ref{fig:tor}, the torque $T$ increases with the depth and the penetration speed. Given a fixed $\omega$, increasing $v$ leads to the decrease of soil discharging capacity, which in turn results in an increase in $P$ and consequently larger $\tau_1$ and $T$.

\section{Steady state analysis}\label{sec:steady}
The above numerical investigations show that the drilling process converges to a steady state, in which the internal and external granular materials flow with constant speeds. In this section, we derive explicitly the relationship between $\zeta$ and $\gamma$ in the steady state. Additionally, we investigate how $\zeta$ is affected by the drill's geometry and the properties of the granular materials.

In the steady state with $\ud u_i/\ud t=0$ and $\ud u_s/\ud t=0$, Eq.\eqref{eq:unidyna} can be re-formulated as
\begin{equation}\label{eq:pressureIO}
\left \{
  \begin{aligned}
 &P = \rho h_i g -\rho u_i^s (v-u_i^s),\\
 &P = \rho h g\sin{\beta}  +  \frac{\rho v u_s^s}{\sin\beta} - \frac{1}{S_o}F_r
  \end{aligned}
  \right.
\end{equation}
where $h_i=m_i/(\rho S_i)$ and $h=l\sin\beta=vt$ represent the sampling height and  the drilling depth at time $t$, respectively. $u_i^s$ and $u_s^s$ represent $u_i$ and $u_s$ in the steady state. By eliminating the term $P$, Eq.\eqref{eq:pressureIO} yields
\begin{equation}\label{eq:prediction3}
  \rho g (h_i-h\sin\beta)-[\rho u_i^s (v-u_i^s)+\frac{\rho v u_s^s}{\sin\beta}]=- \frac{F_r}{S_o}.
\end{equation}

In the steady state, $I$ remains unchanged so that $\mu(I)$ can be replaced by a constant $\overline{\mu}$. Additionally,  figure~\ref{fig:tantheta} shows that $\tan\theta$  is kept unchanged as the drill process enters the steady state. We denote $\theta^s$ as the constant angle. According to Eq.~\eqref{eq:friction}, together with Eqs.\eqref{eq:tau1}, \eqref{eq:sigma2} and \eqref{eq:tau2}, the right term of Eq.\eqref{eq:prediction3} can then be expressed as
\begin{equation}\label{eq:prediction7}
\begin{split}
 -\frac{F_r}{S_o}=k_1\int_0^{l}p(s)\ud s
 &+k_2 p_c l+\mu_0 \rho g l \cos\beta.
\end{split}
\end{equation}
with
\begin{equation}\label{eq:unidyna1def}
\left \{
  \begin{aligned}
 &k_1\equiv\frac{\mu_0+\overline{\mu}\cos\theta^s+\overline{\mu}\mu_0 \sin\theta^s}{\eta}+\frac{2\mu_0}{b\cos\beta},\\
 &k_2\equiv\frac{\overline{\mu} \cos\theta^s+\overline{\mu}\mu_0\sin\theta^s}{\eta}.
  \end{aligned}
  \right.
\end{equation}

According to the pressure distribution $p(s)$ shown in Eq.\eqref{eq:pre_dis},  we get
\begin{equation}\label{eq:integral}
  \int_0^{l}p(s)\ud s=\frac{Pl}{1-\ue^{-\frac{l}{2b}}}-2Pb.
\end{equation}

Finally, combining Eqs.~\eqref{eq:prediction3},\eqref{eq:prediction7} and \eqref{eq:integral} leads to the criteria for the steady state
\begin{equation}\label{eq:prediction8}
  k_1 \frac{Pl}{1-e^{-\frac{l}{2b}}}=G-C,
\end{equation}
where
\begin{equation}\label{eq:unidyna1}
\left \{
  \begin{aligned}
 &G\equiv\rho g (h_i-h\sin\beta)-\mu_0 \rho g l \cos\beta+2k_1Pb-k_2p_cl,\\
 &C\equiv\rho u_i^s (v-u_i^s)+\frac{\rho v u_s^s}{\sin\beta}.
  \end{aligned}
  \right.
\end{equation}

Clearly, coefficients $k_1$, $k_2$ and $C$ are constants in the steady state. We  have $G\sim O(t)$ because $h$, $h_i$, $l$ and $P$ are all $\propto t$. Similar scaling analysis shows that $\frac{Pl}{1-\ue^{-\frac{l}{2b}}} \sim O(t^2)$. For a steady state to be achieved, Eq.~\eqref{eq:prediction8} needs to be always satisfied at large $t$. In another word, it should be valid for all time in the steady state, or on the asymptotic limit of $t$. This condition is true only if $k_1=0$, i.e.
\begin{equation}\label{eq:prediction10}
  k_1\equiv \frac{\mu_0+\overline{\mu}\cos\theta^s+\overline{\mu}\mu_0 \sin\theta^s}{\eta}+\frac{2\mu_0}{b\cos\beta} =0,
\end{equation}
which can be reformulated as
\begin{equation}\label{eq:prediction11}
  \cos\theta^s+\mu_0\sin\theta^s=-\frac{\mu_0}{\overline{\mu}}(\frac{2\eta}{b\cos\beta}+1).
\end{equation}

Eq.~\eqref{eq:prediction11} shows that $\theta^s$ depends on both geometric parameters of the auger drill ($b$ and $\beta$) and properties of granular sample ($\eta$, $\overline{\mu}$ and $\mu_0$). It is irrelevant to the driving conditions of the drilling tool, in agreement with the numerical results shown in figure~\ref{fig:tantheta}.

Equation~\ref{eq:volumeconservation} also suggests a relationship between $u_s^s$ and $u_i^s$:
\begin{equation}\label{eq:prediction12}
  u_s^s=\frac{S_b v-S_i(v-u_i^s)}{S_o}-(\omega \overline{r} \cos\beta+v\sin\beta)
\end{equation}

Noting that $u_\xi =\omega \overline{r} \sin\beta-v\cos\beta$, together with the definition $\tan\theta^s\equiv\frac{u_\xi}{u_s}$, we get,
\begin{equation}\label{eq:prediction13}
  \cot\theta^s =\frac{\frac{S_b}{S_o\cos\beta } -\frac{S_i}{S_o\cos\beta}\frac{v -u_i^s}{v}-(\frac{\omega \overline{r}}{v}+ \tan\beta)}{\frac{\omega \overline{r} \tan\beta}{v}-1}
\end{equation}

We know from Eq.\eqref{eq:inmassrate} that ${\rm d}m_i = \rho \pi r_i^2 (v-u_i) {\rm d}t$. Therefore, the sampling efficiency $\zeta$ becomes
\begin{equation}\label{eq:samplingrate3}
   \zeta= 1-\frac{1}{H}\int_0^H \frac{u_i}{v}\ud h.
\end{equation}

In the steady state with constant $u_i$, Eq.~\eqref{eq:samplingrate3} can be simplified as
\begin{equation}\label{eq:samplingrate2}
   \zeta^s = 1-\frac{u_i^s}{v}.
\end{equation}

Plugging in $\gamma=\frac{\omega b}{2\pi v}=\frac{\omega \overline{r} \tan\beta}{v}$ and $S_b=S_i + S_o/\sin\beta$, Eq.~\eqref{eq:prediction13} can be written as
\begin{equation}\label{eq:prediction14}
  \zeta^s=-\frac{S_o \cos\beta}{S_i}\left(\cot\beta+\cot\theta^s\right)\left(\gamma-1\right)+1\equiv \Gamma
\end{equation}
This equation provides a convenient way to predict analytically sampling efficiency under different conditions. It suggests that $\zeta^s$ decreases linearly as $\gamma$ increases, in agreement with figure~\ref{fig:expcomparison}. Moreover, we find a fixed point $(\gamma=1,~\zeta^s=1$), which is consistent with our aforementioned speculation for $\alpha=\beta$ that it is similar to inserting a straight tube into granular material. In this case, the materials in the auger flight will not be discharged and the sampling efficiency will be 100\%. 

Figure~\ref{fig:invel} indicates that higher penetration velocity leads to a faster convergence of $u_i$ to the steady state value $u_i^s$, and consequently a faster convergence of $\zeta$ to $\zeta^s$ because $\zeta \propto u_i/v$. That means, higher penetration speed, or smaller $\gamma$, yields a better agreement with the analytical prediction of $\zeta$-$\gamma$ relation shown in Eq.~(\ref{eq:prediction14}). Therefore, the $\zeta$-$\gamma$ relation in figure~\ref{fig:expcomparison} becomes closer to the line predicted by Eq.~(\ref{eq:prediction14}) as $\gamma$ decreases. Reciprocally, as $\gamma$ becomes larger, convergence of $\zeta$ takes more time (or larger depth), thus a deviation of $\zeta-\gamma$ curve from a straight line, as illustrated in both experimental and numerical results, can be expected.

\begin{figure} 
\centering
\includegraphics[width=2.8in]{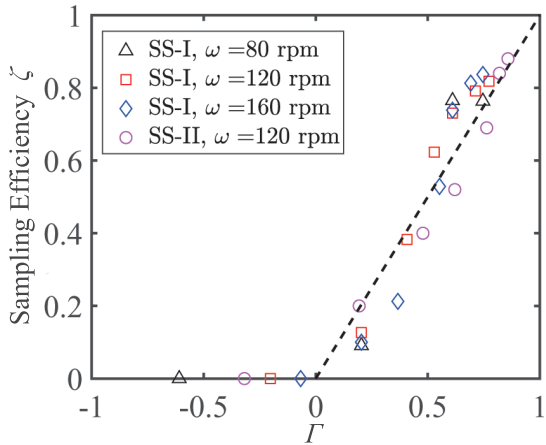}
\caption{The $\zeta^s-\Gamma$ relation of the prediction model in the steady state (dashed line) and the experimental results (markers).}
\label{fig:unify}
\end{figure}

Finally, we compare the analytical model with experimental data directly. The right term of Eq.~(\ref{eq:prediction14}) is denoted by $\Gamma$. $\Gamma$ for the experimental data is computed after parameters presented in Table \ref{tab:exp}. We plot the experimental and numerical results together in figure~\ref{fig:unify} for a direct comparison.

Equation~\ref{eq:prediction14} indicates that the linear relation between $\zeta^s$ and $\gamma$ depends on parameters (e.g., $\mu$, $\mu_0$, $\eta$, $b$, and $r_i$) that are gravity independent. In other words, despite that the evolution of $u_i$ into the steady state is gravity dependent [see Eq. (\ref{eq:unidyna})], sampling efficiency $\zeta^s$ in the steady state is independent of gravitational acceleration, clearly suggesting that auger drilling yields the same sample collection efficiency, regardless of drilling on earth, moon, or other extraterrestrial objects. This can be understood through an analysis on the evolution of gravity dependent terms. For instance, we know that the frictional term in the governing equations increases with drilling depth quadratically, while the gravitational acceleration term is linear with drilling depth. Therefore, as the depth increases, the role of the gravitational acceleration term becomes increasingly weaker, whereas the friction term dominate the dynamics.

\section{Conclusions}\label{sec:conclusion}
To conclude, we experimentally investigate the sampling efficiency of a standard auger drill tool under different drilling conditions and soil properties. The experimental results show that the sampling efficiency decreases monotonically with the growth of the ratio between rotational and penetration speeds. In addition, the sampling efficiency of the drilling tool is found to be sensitive to the soil properties, such as the granular internal friction angle, packing fraction, size distribution, grain scale. We speculate that  the influence of granular friction dominates that arising from other relevant properties, or the influence of granular friction contains that from other soil properties, as previously demonstrated in~\citet{kang2018archimedes}.

In the drilling process, under the cutting effect of the drill bit and the rotation of the drill tool, the soil around the drill tool is fluidized. Therefore, we assume the flowing soil as incompressible fluid and build their governing equations based on the mass conservation and momentum equations of fluid dynamics. Because the flowing soils in the external channel can't be completely discharged to the ground surface, we introduce an effective thickness of the flowing layer $\eta$ that is smaller than the geometrical groove depth $a$. The numerical solutions of the governing equations show the same features as the experimental results: the sampling efficiency decreases with the ratio of rotation to penetration speed and doesn't depend their specific values. By setting a smaller $\eta$ to the Soil-I with higher granular friction than Soil-II, the theoretical results of both types of soils can agree well with the experimental results.

Note that the numerical solutions of the theoretical model converge as drilling deepens, leading to a steady state under which the governing equations can be solved analytically. Consequently, we analyze the theoretical model in the steady state and find that the ratio of the normal and streamwise velocity components of granular flow in the external channel is a constant being independent on the penetration speed, explaining why the sampling efficiency is independent on $v$ and $\omega$ individually. Moreover, we obtain an analytical prediction for the sampling efficiency. The outcome shows that the sampling efficiency decreases linearly with the speed ratio $\gamma$ and it goes through $(\gamma,\zeta_s)=(1,1)$, in consistent with the qualitative analysis in \S~\ref{sec:experiment}. Importantly, this prediction shows that the sampling efficiency is irrelevant to gravity in the steady state, which has important significance for widespread applications. Note that gravity indeed plays a role before the steady state is established.

Future work will be devoted to using DEM simulations for further analysis on granular flows in both channels, particularly the pathway towards steady states, along with a characterization of the pressure distribution in both channels to assist in further development of the model, as well as in guiding verification experiments. In addition, it is also important to investigate granular flow profile in the helical external channel during the drilling process within the framework of $\mu - I$ rheology and explore the possibility of instabilities induced by shear flow \cite{Cortet2009, Boerzsoenyi2009, Brodu2013}.

\section{Acknowledgement}\label{sec:ack}
This work has been supported by the National Natural Science Foundation of China (NSFC: 11932001). KH acknowledges the Startup Grant from Duke Kunshan University. We thank Rafi Blumenfeld for fruitful discussions and the help of Xiaoming Lai from Beijing Spacecrafts, China Academy of Space Technology.

\section{Declaration of interests}
The authors report no conflicts of interest.

\bibliographystyle{jfm}
\bibliography{Drilling}

\end{document}